\documentstyle[11pt,newpasp,twoside,epsf]{article}
\def\edcomment#1{\iffalse\marginpar{\raggedright\sl#1\/}\else\relax\fi}
\reversemarginpar

\begin{document}
\title{The optical/near-IR spectral energy distribution of the GRB~000210 host galaxy}
\author{J.   Gorosabel$^1$,   L.   Christensen$^2$,   J.   Hjorth$^3$,
J.U. Fynbo$^4$,  H. Pedersen$^3$, B.L.  Jensen$^3$, M.I. Andersen$^5$,
N. Lund$^6$,  A.O. Jaunsen$^7$,  J.M. Castro Cer\'on$^8$,  A.J. Castro
Tirado$^1$,  A.  Fruchter$^9$,   J.  Greiner$^{10}$,  E.  Pian$^{11}$, P.M.
Vreeswijk$^7$,    I.    Burud$^9$,   F.    Frontera$^{12,13}$, L.
Kaper$^{14}$,    S.    Klose$^{15}$,   C.    Kouveliotou$^{16}$, N.
Masetti$^{13}$,  E.  Palazzi$^{13}$,  J.  Rhoads$^9$,  E.  Rol$^7$,
I. Salamanca$^7$, Tanvir$^{17}$, R.A.M.J. Wijers$^{14}$ \& E.P.J. van den
Heuvel$^{14}$}
\affil{$^1$IAA-CSIC;   $^2$AIP  Potsdam;   $^3$Univ.   of  Copenhagen; $^4$Univ. of Aarhus; $^5$Univ. of Oulu; $^6$DSRI, Copenhagen; $^7$ESO, Santiago;  $^8$ROA,   C\'adiz;  $^9$STScI,  Baltimore;   $^{10}$MPE, Garching;  $^{11}$INAF-  Astron.  Obs.  of  Trieste;  $^{12}$Univ.  of Ferrara; $^{13}$IASF/CNR, Bologna;  $^{14}$Univ. of Amsterdam; $^{15}$TLS,  Tautenburg;  $^{16}$NASA/MSFC,   Huntsville;  $^{17}$Univ.  of Hertfordshire}

\begin{abstract}

  We  report on  UBVRIZJsHKs-band photometry  of the  dark GRB  000210 host
galaxy.   Fitting a  grid  of  spectral templates  to  its Spectral  Energy
Distribution     (SED),    we     derived     a    photometric     redshift
($z=0.842^{+0.054}_{-0.042}$)  which  is in  excellent  agreement with  the
spectroscopic one ($z=0.8463\pm  0.0002$; Piro et al. 2002).   The best fit
to  the SED  is obtained  with a  blue starburst  template with  an  age of
$0.181^{+0.037}_{-0.026}$ Gyr.  We discuss the implications of the inferred
low value of $A_{\rm V}$ and the age of the dominant stellar population for
the non detection of the GRB~000210 optical afterglow.


\end{abstract}

\section{Introduction}

 GRB~000210 is currently one of the few systems that allow a detailed study
of  a galaxy  which hosted  a dark  GRB.  The  burst exhibited  the highest
$\gamma$-ray  peak flux  among the  GRBs  localized during  the entire  SAX
operation  (Piro et  al.  2002).   However, no  optical afterglow  (OA) was
detected  in spite  of a  deep search  carried out  $\sim16$ hrs  after the
gamma-ray event  (Gorosabel et  al.  2000a).  X-ray  observations performed
with the Chandra  X-ray telescope 21 hrs after the  GRB localised the X-ray
afterglow of  the burst to an  accuracy of $0\farcs6$ (Piro  et al.  2002).
The optical search revealed an extended constant source coincident with the
X-ray afterglow which was proposed as the GRB host galaxy (Gorosabel et al.
2000b).   Based on the  detection of  a single  host galaxy  spectral line,
interpreted  as  [\ion{O}{II}] $\lambda$  3727  \AA,  Piro  et al.   (2002)
proposed  a redshift  of $z=0.8463  \pm  0.0002$.  Recently  Berger et  al.
(2002) and  Barnard  et  al.    (2003)  have  reported  $\sim$2.5  $\sigma$
detections of sub-mm emission towards  the position of the GRB~000210 host,
suggesting  a Star  Formation  Rate (SFR)  of  several hundred  $M_{\odot}$
yr$^{-1}$.

By fitting SED  templates to the near-IR/optical photometric  points we can
estimate the  SFR, the  extinction ($A_{\rm V}$)  and the  dominant stellar
population  age.  The  inferred  stellar  age can  help  us to  distinguish
between the two families of GRB progenitors; the hypernovae (stellar age $<
10^{8}$ yr)  and the  binary compact mergers  (stellar age $>  10^{8}$ yr).
The  SED  also   provides  information  on  the  extinction   law  and  the
metallicity.   Furthermore,  the   SED  fitting  provides  the  photometric
redshift of the  host.  The comparison of the  derived photometric redshift
with  the  spectroscopic one  allows  us  to  eliminate incorrect  SED  fit
solutions and to check the consistency of our method.  In the present paper
we present the most intensive multi-colour host galaxy imaging performed to
date  and  the  results obtained  when  SED  templates  are fitted  to  the
photometric points.

\section{Method: constructing SED templates}
\label{method}

A number  of optical/near-IR resources  (VLT, NTT, 1.54-m Danish  and 3.6-m
ESO telescopes) have been used in order to compose a well sampled SED, from
the U  to the Ks  band. The fits  of the SEDs  have been carried  out using
Hyperz\footnote{  http://webast.ast.obs-mip.fr/hyperz/} (Bolzonella  et al.
2000).   Eight synthetic  spectral types  were used  representing Starburst
galaxies (Stb), Ellipticals (E), Lenticulars  (S0), Spirals (Sa, Sb, Sc and
Sd) and  Irregular galaxies (Im).   The time evolution  of the SFR  for all
galaxy types  is represented  by an exponential  model, i.e.   SFR $\propto
\exp(-t/\tau)$, ranging $\tau$ from 0 (Stb)  to 30 Gyr (Sd).  The SFR of Im
galaxies are represented by a constant SFR ($\tau \rightarrow \infty$).

\begin{table}
\begin{center}
\caption{\label{table1} The table displays  the parameters of the best
host galaxy SED fit when  several IMFs, indicated in the first column,
are adopted.   The second column  provides the confidence of  the best
fit (given  by $\chi^{2}/$dof).   The derived photometric  redshift is
displayed in  the third  column (and the  corresponding 68\%  and 99\%
percentile  errors).  In  the fourth  and fifth  columns  the template
family  of the  best fitted  SED and  the stellar  population  age are
given.  The  sixth column displays the host  galaxy extinction $A_{\rm
V}$.  The absolute rest frame B-band magnitude, $M_B$, is given in the
last column.}

\vspace{0.2cm}
\begin{tabular}{@{}lcccccc@{}}
\hline
 IMF     & $\chi^{2}/$dof& Photometric redshift & Template &  Age  & $A_{\rm V}$& $M_B$ \\

         &           &$z^{+ p68\%, p99\%}_{- p68\%, p99\%}$&& (Gyr)&            &      \\
\hline                                                                                
 Sa55    & $1.096$   &$0.842^{+0.054, 0.158}_{-0.042, 0.279}$& Stb &0.181& 0.00 &-20.16\\
         &           &                                       &           &      &      \\
 MiSc79  & $1.046$   &$0.836^{+0.087, 0.140}_{-0.053, 0.244}$& Stb &0.181& 0.00 &-20.16\\
         &           &                                       &           &      &      \\
 Sc86    & $0.903$   &$0.757^{+0.067, 0.219}_{-0.044, 0.132}$& S0  &1.015& 0.00 &-19.90\\
\hline
\end{tabular}
\end{center}
\end{table}

Once the population of stars is generated following the time evolution
given by the assigned SFR, the  mass of the newly formed population is
distributed in stars following an assumed Initial Mass Function (IMF).
Three  IMFs have  been considered:  Miller \&  Scalo  (1979), Salpeter
(1955), and Scalo (1986).  These IMFs will be abbreviated hereafter as
MiSc79, Sa55  and Sc86, respectively.  

Several extinction laws have been tested for the SED fitting, yielding
all of them  very similar results.  Thus hereafter  the extinction law
given  by Calzetti  et  al.   (2000) for  starburst  galaxies will  be
adopted.  A  more detailed  description on the  impact of  the assumed
extinction laws and IMFs can be found at Gorosabel et al. (2003).

\section{Results and conclusions}

 The inferred  photometric redshifts  are (independently of  the extinction
 law and IMF  assumed) consistent, within the 99\%  percentile error range,
 with the spectroscopic  redshift (see Table 1).  However,  within the 68\%
 precentile ($\sim 1\sigma$) error range, only the Sa55 and MiSc79 IMFs are
 consistent, Sc86 is not.  Thus, we conclude that the fit based on the Sc86
 IMF (third  row of Table  1) provides the  least likely fit  solution.  As
 shown in the  first two rows of the table, the  resolution of our template
 grid is  not able  to make  a distinction between  most of  the properties
 (Age, $A_{\rm V}$, $M_B$) derived for the Sa55 and MiSc79 IMFs.

 Hence, the  SED of  the host  galaxy is consistent  with a  blue starburst
 template with an age of $\sim$0.181 Gyr and a very low extinction ($A_{\rm
 V} \sim 0$, see Fig. 1).   Based on the restframe UV flux (Kennicutt 1998)
 a star formation  rate of $2.1 \pm 0.2  M_{\odot}$ yr$^{-1}$ is estimated.
 The absolute  restframe B-band  magnitude of the  host is $M_B  = -20.16$,
 which corresponds to  $L = 0.5\pm0.2 L^{\star}$, depending  on the assumed
 value of $M_B^{\star}$ (Lilly et al.  1995; Schechter 1976). Therefore, we
 conclude that the GRB 000210 host is very likely a subluminous galaxy.

\begin{figure}
\plottwo{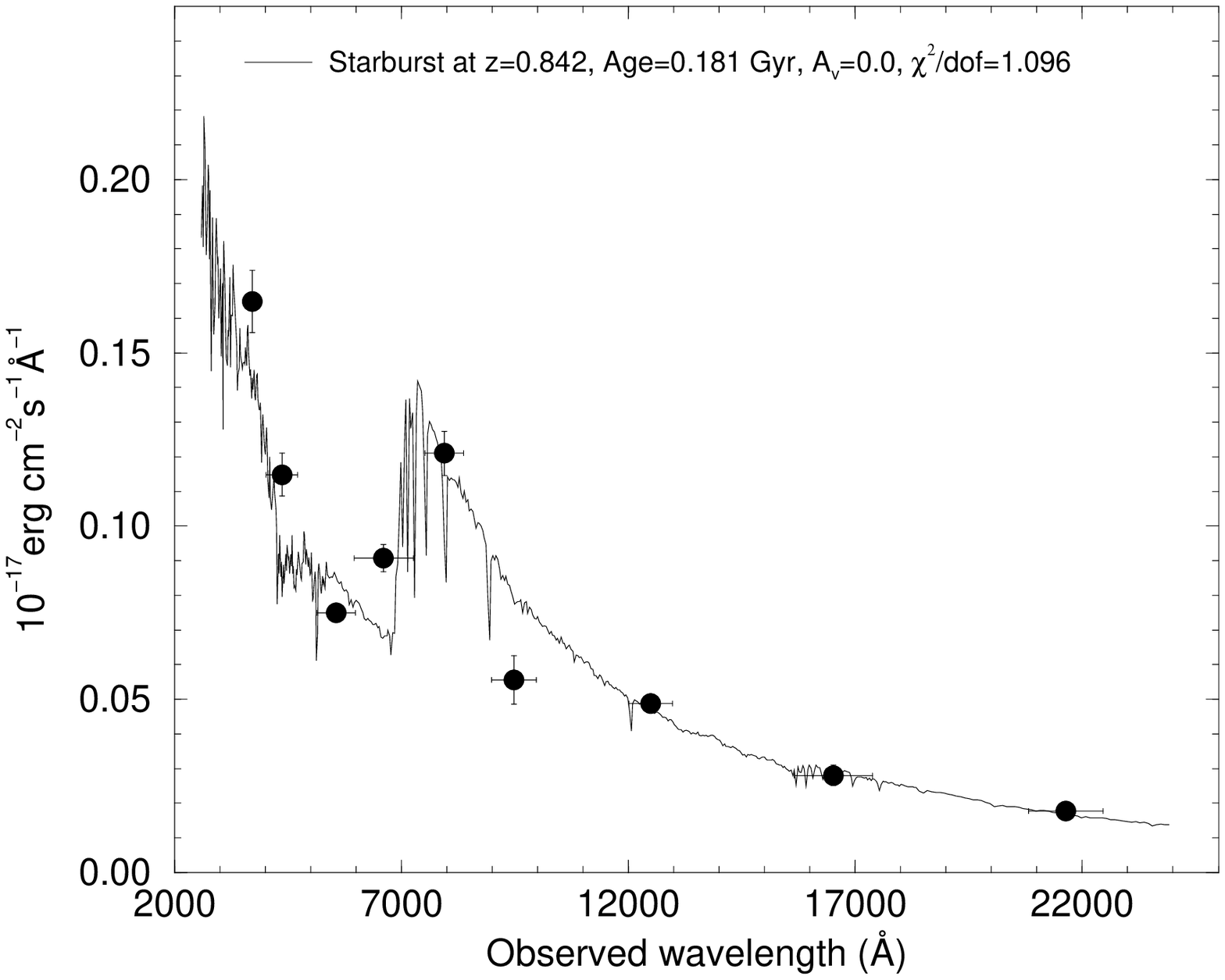}{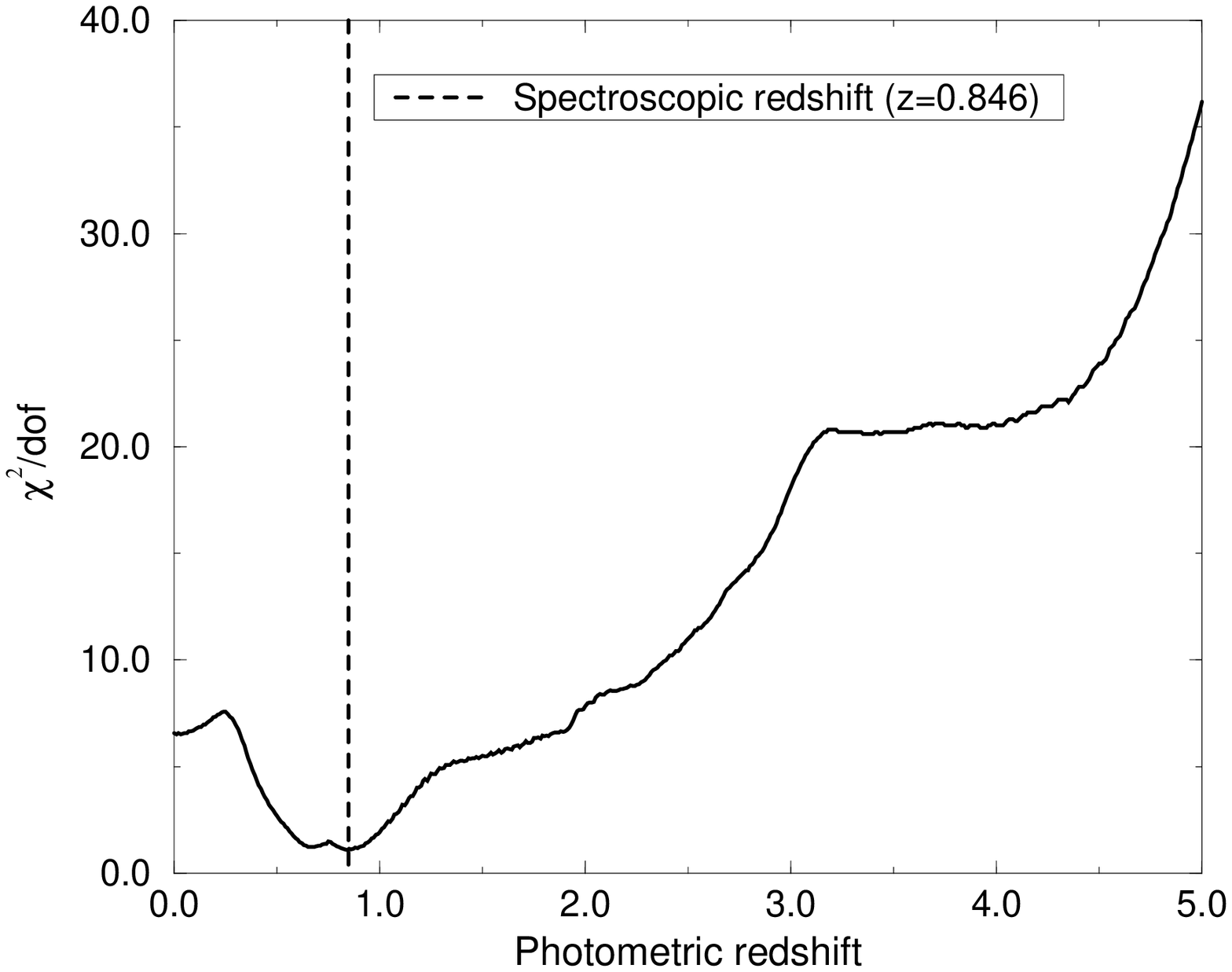}
\caption{\label{fig4}  {\em Left:}  The figure  shows  the UBVRIZJsHKs-band
photometric  points  of  the  GRB~000210  host  galaxy.   The  solid  curve
represent the best fitted SED  ($\chi^2$/dof = $1.096$), corresponding to a
starburst synthetic  template at  a redshift of  $z = 0.842$.   The derived
value  of  the  starburst age  corresponds  to  $0.181$  Gyr.  The  fit  is
consistent with a very low  host galaxy global extinction. {\em Right:} The
plot displays the evolution of the fitted SED $\chi^2/$dof as a function of
the  photometric   redshift.   The  dotted  vertical   line  indicates  the
spectroscopic  redshift proposed  by Piro  et  al.  (2002).   As shown  the
minimum of $\chi^2/$dof (at $z=0.842$) is consistent with the spectroscopic
redshift.  }
\end{figure}

 The low value  of the extinction obtained in the SED  fit ($A_{\rm V} \sim
 0$) makes  it difficult to explain  the optical darkness  of GRB~000210 in
 terms of the global host galaxy dust extinction. If dust extinction is the
 reason of  the lack  of optical afterglow  emission, then  the circumburst
 region has to  be very compact and localised  around the progenitor.  This
 clumpy and  fragmented ISM  may help to  explain the  apparent discrepancy
 between our  SFR estimate (derived from  the galaxy UV flux,  $2.1 \pm 0.2
 M_{\odot}$  yr$^{-1}$  )  and  the  one recently  reported  based  on  the
 sub-millimeter range (Several hundred $M_{\odot}$ yr$^{-1}$, Berger et al.
 2002).

 Both the collapsar  and the binary merging models  show severe limitations
 to explain the visible stellar age and the line of sight \ion{H}{I} column
 density  (derived  from the  afterglow  X-ray  spectrum) respectively.   A
 solution to this problem would be the existence of a younger population of
 stars  (several Myr of  age) hidden  by the  clumpy ISM.   Such population
 (which  would include  the GRB  progenitor)  would not  have a  detectable
 impact on the host galaxy SED probed by our observations.

\vfill
\eject

\end{document}